# On the space complexity of one-pass compression

Travis Gagie

Department of Computer Science
University of Toronto
travis@cs.toronto.edu

STUDENT PAPER

**Abstract.** We study how much memory one-pass compression algorithms need to compete with the best multi-pass algorithms. We call a one-pass algorithm an $f(n, \ell)$-footprint compressor if, given $n$, $\ell$ and an $n$-ary string $S$, it stores $S$ in $\bigl(O(H_\ell(S)) + o(\log n)\bigr) |S| + O(n^{\ell+1} \log n)$ bits — where $H_\ell(S)$ is the $\ell$th-order empirical entropy of $S$ — while using at most $f(n, \ell)$ bits of memory. We prove that, for any $\epsilon > 0$ and some $f(n, \ell) \in O(n^{\ell+\epsilon} \log n)$, there is an $f(n, \ell)$-footprint compressor; on the other hand, there is no $f(n, \ell)$-footprint compressor for $f(n, \ell) \in o(n^\ell \log n)$.

## 1 Introduction

One-pass compression has been studied extensively and many of the best compression algorithms — e.g., LZ77, LZ78 and PPM — are one-pass algorithms. It is useful when compressing files too large to fit in internal memory, since each page need be brought from external memory only once; it is crucial when compressing data that are time-sensitive and must be tranmitted as they are generated, or from data streams (see, e.g., [1]) — sources that produce so many data they cannot all be stored and must be manipulated as they pass. Even one-pass algorithms have a disadvantage, though: most build and store models and, in general, the better the compression, the bigger the model. There are several techniques for limiting one-pass algorithms' memory consumption, such as using a sliding window over the data, but it is not always clear how these affect compression. In this paper we prove worst-case bounds on how much memory one-pass algorithms need to compete with the best multi-pass algorithms.

The first notable work on one-pass compression was done by Faller [8] in 1973 and Gallager [9] in 1978, toward a dynamic version of Huffman's algorithm [10]. In 1985 Knuth [12] used their results in an algorithm that takes time linear in the size of its output. It was the basis of the Unix utility compact and is sometimes known as the FGK algorithm — the authors' initials — to distinguish it from an improved version of dynamic Huffman coding that Vitter [18] published in 1987. In the meantime, Ziv and Lempel had invented their well-known algorithms LZ77 [19] and LZ78 [20], which were the bases of, e.g., the Unix utility compress, gzip, WinZip® and, more recently, 7-Zip. Dynamic arithmetic coding was developed in the 1980s (see, e.g., [15]) and, generalizing it to use context, PPM (see, e.g., [5]); although PPM is somewhat slower and less space efficient than other popular algorithms, it is generally considered to achieve the best compression. Finally, Bentley, Sleator, Tarjan and

Wei [2] and Elias [7] independently invented move-to-front compression, which is especially intersting for us because its space complexity does not depend on its input's length; we use this property in Section 3 when proving our upper bound.

Algorithms based on the Burrows-Wheeler Transform [3] represent the state of the art in multi-pass compression — e.g., bzip2. Kaplin, Landau and Verbin [11] recently proved one such algorithm, given an $n$-ary string $S$, stores $S$ in $\left(cH_\ell(S) + \log \zeta(c) + r\right)|S| + cn^{\ell+1}\log n$ bits, for any $c > 1$ and $\ell \geq 0$. By $H_\ell(S)$ we mean the $\ell$th-order empirical entropy of $S$; log means $\log_2$; $\zeta$ is the Riemann zeta function; and $r$ is the redundancy of a simpler compression algorithm (e.g., arithmetic coding or Huffman coding) chosen as a subroutine. Their analysis was based on an earlier one by Manzini [14], which established empirical entropy as a popular complexity metric. Since even Manzini's analysis is relatively recent, we discuss empirical entropy in Section 2. To compare one-pass and multi-pass algorithms theoretically, we use Kaplan, Landau and Verbin's analysis as a benchmark. We call a one-pass algorithm an $f(n,\ell)$-footprint compressor if, given $n$, $\ell$ and an $n$-ary string $S$, it stores $S$ in $\left(O(H_\ell(S)) + o(\log n)\right)|S| + O(n^{\ell+1}\log n)$ bits while using at most $f(n,\ell)$ bits of memory. In Section 3 we prove nearly tight bounds on footprints: for any $\epsilon > 0$ and some $f(n,\ell) \in O(n^{\ell+\epsilon}\log n)$, there is an $f(n,\ell)$-footprint compressor; for $f(n,\ell) \in o(n^\ell \log n)$, however, there is no $f(n,\ell)$-footprint compressor.

## 2 Empirical entropy

Markov processes have long been the most popular models for many kinds of data; e.g., Shannon [17] fitted zeroth-, first- and second-order Markov processes to English, gave samples of their output and wrote "the resemblance to ordinary English text increases quite noticeably at each of the above steps", and "a sufficiently complex stochastic process will give a statisfactory representation" of natural written language. Of course, there is a important difference between representation and equivalence: Chomsky [4] argued a probabilistic model cannot determine whether a novel sentence is grammatical (e.g., "Colorless green ideas sleep furiously.") or not (e.g., "Furiously sleep ideas green colorless.") and concluded "the notion 'grammatical in English' cannot be identified in any way with the notion 'high order of statistical approximation to English'"; in other words, people are not Markov processes.

The self-information of a string $S$ with respect to a source $M$ — i.e., the negative logarithm of the probability $M$ generates $S$ — measures how many bits we need to store $S$ when using $M$ as a model. The $\ell$th-order empirical entropy $H_\ell(S)$ of $S$ is $1/|S|$ times the minimum self-information of $S$ with respect to an $\ell$th-order Markov process. Thus, comparing $H_\ell(S)$ to the logarithm of the alphabet size tells us how much we can compress $S$ when using an $\ell$th-order Markov process as a model — regardless of how $S$ is generated. As Manzini [14] wrote, "the empirical entropy resembles the entropy defined in the probabilistic setting (for example, when the input comes from a Markov source) [but] is defined for any string and can be used to measure the performance of compression algorithms without any assumption on the input."

Another way to view the $\ell$th-order empirical entropy of $S$ is as our expected uncertainty about a randomly chosen character, given a context of length $\ell$. Let $s_i$ denote the $i$th character of $S$ and consider the following experiment: $i$ is chosen uniformly at random from $\{1, \ldots, |S|\}$; if $i \leq \ell$, then we are told $s_i$; otherwise, we are told $s_{i-\ell} \cdots s_{i-1}$. Our expected uncertainty about the random variable $s_i$ — its expected entropy — is

$$H_\ell(S) = \begin{cases} \displaystyle\sum_{a \in S} \frac{\#_a(S)}{|S|} \log \frac{|S|}{\#_a(S)} & \text{if } \ell = 0, \\ \displaystyle\frac{1}{|S|} \sum_{|\alpha|=\ell} |S_\alpha| \cdot H_0(S_\alpha) & \text{if } \ell \geq 1. \end{cases}$$

Here, $a \in S$ means character $a$ occurs in $S$; $\#_a(S)$ is the number of occurrences of $a$ in $S$; and $S_\alpha$ is the string obtained by concatenating the characters immediately following occurrences of string $\alpha$ in $S$ — the length of $S_\alpha$ is the number of occurrences of $\alpha$ in $S$ unless $\alpha$ is a suffix of $S$, in which case it is 1 less. Notice that, if $n$ is the alphabet's size, then $H_{\ell+1}(S) \leq H_\ell(S) \leq \log n$ for all $\ell$. For example, if $S$ is the string TORONTO, then

$$H_0(S) = \frac{1}{7} \log 7 + \frac{3}{7} \log \frac{7}{3} + \frac{1}{7} \log 7 + \frac{2}{7} \log \frac{7}{2} \approx 1.84 ,$$

$$H_1(S) = \frac{1}{7} \Big( H_0(S_\text{N}) + 2H_0(S_\text{O}) + H_0(S_\text{R}) + 2H_0(S_\text{T}) \Big)$$
$$= \frac{1}{7} \Big( H_0(\text{T}) + 2H_0(\text{RN}) + H_0(\text{O}) + 2H_0(\text{OO}) \Big)$$
$$= 2/7 \approx 0.29$$

and all higher-order empirical entropies of $S$ are 0. This means if someone chooses a character uniformly at random from TORONTO and asks us to guess it, then our uncertainty is about 1.84 bits. If they tell us the preceding character before we guess, then on average our uncertainty is about 0.29 bits; if they tell us the preceding two characters, then we are certain of the answer.

## 3 Upper and lower bounds on footprints

Our upper bound is based on a generalization of move-to-front compression (MTF). To compress an $n$-ary string $S$, MTF starts with a list storing the numbers from 0 to $n-1$; for each character $s_i$ in $S$, it prints $s_i$'s position in the list, encoded with Elias' delta code [6], then moves $s_i$ to the front of the list. Despite its simplicity, MTF is quite efficient: it stores $S$ in $\big(H_0(S) + 2\log(H_0(S) + 1) + 1\big)|S| + O(n \log n)$ bits while using $O(n \log n)$ bits of memory.

**Theorem 1.** *For any $\epsilon > 0$ and some $f(n, \ell) \in O(n^{\ell+\epsilon} \log n)$, there is an $f(n, \ell)$-footprint compressor.*

*Proof.* Consider the algorithm $\mathcal{A}_{n,\ell}$ that, given an $n$-ary string $S$, keeps a list $Q_\alpha$ of maximum size $\lfloor n^\epsilon \rfloor$ for each possible $\ell$-tuple $\alpha$. For $1 \leq i \leq \ell$, $\mathcal{A}_{n,\ell}$ prints the $\lceil \log n \rceil$-bit binary representation of the $i$th character $s_i$ of $S$. For $i \geq \ell + 1$, if $s_i$ is stored in $Q_{s_{i-\ell},\ldots,s_{i-1}}$, then $\mathcal{A}_{n,\ell}$ prints 1 followed by $s_i$'s position in $Q_{s_{i-\ell},\ldots,s_{i-1}}$ — encoded with the delta code — and moves $s_i$ to the front; otherwise, $\mathcal{A}_{n,\ell}$ prints 0 followed by the $\lceil \log n \rceil$-bit binary representation of $s_i$, then inserts $s_i$ at the front and, if necessary, deletes the last character in $Q_{s_{i-\ell},\ldots,s_{i-1}}$.

It takes $O(n^{\ell+\epsilon} \log n)$ bits to store all $n^\ell$ lists. After $\mathcal{A}_{n,\ell}$ has printed $s_1,\ldots,s_\ell$ in binary, it becomes equivalent to $n^\ell$ copies of $\mathcal{A}_{n,0}$: for each possible $\ell$-tuple $\alpha$, a copy of $\mathcal{A}_{n,0}$ operates on the string $S_\alpha$ obtained by concatenating the characters immediately following occurrences of $\alpha$ in $S$. Thus, by the definition of empirical entropy, if $\mathcal{A}_{n,0}$ prints $\bigl(O(H_0(S_\alpha)) + o(\log n)\bigr) |S_\alpha| + O(n \log n)$ bits when given any $S_\alpha$, then $\mathcal{A}_{n,\ell}$ prints

$$\sum_{|\alpha|=\ell} \bigl((H_0(S_\alpha)) + o(\log n)\bigr) |S_\alpha| + O(n^{\ell+1} \log n)$$
$$= \bigl(O(H_\ell(S)) + o(\log n)\bigr) |S| + O(n^{\ell+1} \log n)$$

bits while using $O(n^{\ell+\epsilon} \log n)$ bits of memory — i.e., $\mathcal{A}_{n,\ell}$ is an $f(n,\ell)$-footprint compressor for some $f(n,\ell) \in O(n^{\ell+\epsilon} \log n)$.

If $\epsilon \geq 1$, then $\mathcal{A}_{n,0}$ behaves like MTF, except that it prepends a 1 to each codeword. Suppose $\epsilon < 1$, $\mathcal{A}_{n,0}$ is given $S_\alpha$ and that, when processing some character $s_i$, MTF would print the encoding of $x$ in the delta code. If $x \leq \lfloor n^\epsilon \rfloor$, then $\mathcal{A}_{n,0}$ prints a 1 followed by the same codeword; if $x > \lfloor n^\epsilon \rfloor$ — in which case the codeword for $x$ in the delta code is at least $\epsilon \log n$ bits — then $\mathcal{A}_{n,0}$ prints a 0 followed by a $\lceil \log n \rceil$-bit number. In either case, $\mathcal{A}_{n,0}$ prints at most $1/\epsilon$ times the number of bits MTF would print, plus 2. Thus, $\mathcal{A}_{n,0}$ stores $S_\alpha$ in $\frac{1}{\epsilon} \bigl(H_0(S) + 2\log(H_0(S_\alpha) + 1) + 3\bigr) |S_\alpha| + O(n \log n) \in \bigl(O(H_0(S_\alpha)) + O(\log \log n)\bigr) |S_\alpha| + O(n \log n)$ bits. □

Our lower bound is based on arguments about objects' Kolmogorov complexities [13]. The Kolmogorov complexity $K(X)$ of an object $X$ is the minimum space needed to store $X$ — more formally, the length in bits of the shortest program that returns $X$; the choice of any Turing-equivalent programming language does not affect $K(X)$ by more than a constant term. By a simple diagonalization, Kolmogorov complexity is neither computable nor even approximable, so it is most often used for proving lower bounds. In our proof, we use three properties of Kolmogorov complexity: the Kolmogorov complexity of any fixed, finite object is a constant; if an object can be computed from other objects, then its Kolmogorov complexity is at most the sum of theirs plus a constant; and, in any set of size $k$, nearly all elements have Kolmogorov complexity $\Omega(\log k)$.

**Theorem 2.** *For $f(n,\ell) \in o(n^\ell \log n)$, there is no $f(n,\ell)$-footprint compressor.*

*Proof.* Assume there is an $f(n,\ell)$-footprint compressor $\mathcal{A}$. As a one-pass algorithm, $\mathcal{A}$'s future behaviour is determined only by its $f(n,\ell)$-bit memory and its unread input; thus, the function $\mathcal{A}(n, \ell, \cdot)$ is computed by some $2^{f(n,\ell)}$-state transducer. Let $T_{n,\ell}$ be the lexicograpically first such transducer and let $\delta^*$ be its extended transition relation. We can construct $T_{n,\ell}$ from $\mathcal{A}$, $n$ and $\ell$, so $K(T_{n,\ell}) \in o(n^\ell \log n)$.

Let $S_{n,\ell}$ be the first $n^\ell$ characters of any $n$-ary linear sequence de Bruijn sequence of order $\ell$. An $n$-ary linear de Bruijn sequence of order $\ell$ is an $n$-ary string containing each possible $\ell$-tuple exactly once; it has length $n^\ell + \ell - 1$ and its first and last $\ell - 1$ characters are the same. Notice $H_\ell(S_{n,\ell}^k) = 0$ for any positive $k$. Let $B$ be the shortest binary string such that there are reachable states $q_1$ and $q_2$ in $T_{n,\ell}$ with $(q_1, S_{n,\ell}, B, q_2) \in \delta^*$; i.e., if $T_{n,\ell}$ is state $q_1$ and reads $S_{n,\ell}$, then it prints $B$ and stops in $q_2$. By the definition of a compressor, $\left|\mathcal{A}\left(n, \ell, S_{n,\ell}^{n^2}\right)\right| \in o\left(\left|S_{n,\ell}^{n^2}\right| \log n\right) = o(n^{\ell+2} \log n)$, so $|B| \in o(n^\ell \log n)$.

There cannot be another string $S'_{n,\ell}$ such that $(q_1, S'_{n,\ell}, B, q_2) \in \delta^*$; otherwise, for some $n$-ary prefix $P$, we would have $\mathcal{A}(n, \ell, PS'_{n,\ell}) = \mathcal{A}(n, \ell, PS_{n,\ell})$ and $\mathcal{A}$ would not be lossless. Thus, we can construct $S_{n,\ell}$ from $T_{n,\ell}$, $q_1$, $q_2$ and $B$, so $K(S_{n,\ell}) \leq 2 \log |T_{n,\ell}| + o(n^\ell \log n) = 2f(n,\ell) + o(n^\ell \log n)$. On the other hand, there are $(n!)^{n^{\ell-1}}$ $n$-ary linear de Bruijn sequences of order $\ell$ [16], so $K(S_{n,\ell}) \in \Omega(n^\ell \log n)$ for nearly all choices of $S_{n,\ell}$. Therefore, $f(n,\ell) \in \Omega(n^\ell \log n)$. □

## Acknowledgments

Many thanks to Giovanni Manzini and Charles Rackoff for their supervision, and to Paolo Ferragina and Roberto Grossi for their advice.

## References

1. B. Babcock, S. Babu, M. Datar, R. Motwani, and J. Widom. Models and issues in data stream systems. In *Proceedings of the 21st Symposium on Principles of Database Systems*, pages 1–16, 2002.
2. J. L. Bentley, D. D. Sleator, R. E. Tarjan, and V. K. Wei. A locally adaptive data compression scheme. *Communications of the ACM*, 29:320–330, 1986.
3. M. Burrows and D.J. Wheeler. A block-sorting lossless data compression algorithm. Technical Report 24, Digital Equipment Corporation, 1994.
4. A. N. Chomsky. *Syntactic Structures*. Mouton, 1957.
5. J. G. Cleary and W. J. Teahan. Unbounded length contexts for PPM. *The Computer Journal*, 40:67–75, 1997.
6. P. Elias. Universal codeword sets and representations of the integers. *IEEE Transactions on Information Theory*, 21:194–203, 1975.
7. P. Elias. Interval and recency rank source coding: Two on-line adaptive variable length schemes. *IEEE Transactions on Information Theory*, 33:3–10, 1987.
8. N. Faller. An adaptive system for data compression. In *Record of the 7th Asilomar Conference on Circuits, Systems and Computers*, pages 593–597, 1973.
9. R. G. Gallager. Variations on a theme by Huffman. *IEEE Transactions on Information Theory*, 24:668–674, 1978.
10. D. A. Huffman. A method for the construction of minimum-redundancy codes. *Proceedings of the IRE*, 40:1098–1101, 1952.
11. H. Kaplan, S. Landau, and E. Verbin. A simpler analysis of Burrows-Wheeler based compression. In *Proceedings of the 17th Symposium on Combinatorial Pattern Matching*, pages 282–293, 2006.
12. D. E. Knuth. Dynamic Huffman coding. *Journal of Algorithms*, 6:163–180, 1985.
13. M. Li and P. Vitányi. *An Introduction to Kolmogorov Complexity and Its Applications*. Springer-Verlag, 2nd edition, 1997.
14. G. Manzini. An analysis of the Burrows-Wheeler Transform. *Journal of the ACM*, 48:407–430, 2001.

15. A. Moffat, R. M. Neal, and I. H. Witten. Arithmetic coding revisited. *ACM Transactions on Information Systems*, 16:256–294, 1998.
16. V. R. Rosenfeld. Enumerating De Bruijn sequences. *MATCH Communications in Mathematical and in Computer Chemistry*, 45:71–83, 2002.
17. C. E. Shannon. A mathematical theory of communication. *Bell System Technical Journal*, 27:379–423, 623–656, 1948.
18. J. S. Vitter. Design and analysis of dynamic Huffman codes. *Journal of the ACM*, 34:825–845, 1987.
19. J. Ziv and A. Lempel. A universal algorithm for sequential data compression. *IEEE Transactions on Information Theory*, 23:337–343, 1977.
20. J. Ziv and A. Lempel. Compression of individual sequences via variable-rate coding. *IEEE Transactions on Information Theory*, 24:530–536, 1978.